\documentstyle[
aps]{revtex}

\begin{document}

\title{Aggregation number distributions and mesoglobules in\\
dilute solutions of diblock and triblock copolymers}


\author{E.G.~Timoshenko\thanks{Corresponding author.
Web page: http://darkstar.ucd.ie; 
E-mail: Edward.Timoshenko@ucd.ie} \and
Yu.A.~Kuznetsov}
\address{
Theory and Computation Group,
Department of Chemistry, University College Dublin,
Belfield, Dublin 4, Ireland
}
\date{\today}

\maketitle

\begin{abstract}
We investigate the aggregation number and size distributions 
for inter--molecular clusters of amphiphilic diblock and triblock 
copolymers in poor solvent at very low concentrations.
Diblocks and triblocks with hydrophilic ends are shown to possess 
narrow distributions corresponding to formation of monodispersed mesoglobules.
Diblocks with hydrophobic ends are found to produce inter--cluster 
multimers due to bridging by the hydrophilic middle blocks, resulting 
in polydisperse distributions.
Implications of these observations for preparation of 
monodispersed nanoparticles and, potentially, understanding of the 
quaternary structure of proteins are discussed.
\end{abstract}


\section{Introduction}

The phenomena of coil--to--globule transition for a single polymer molecule
and phase separation in polymer solutions are foundations
of the modern polymer science \cite{Polymer-books,Experiment}.
In recent years there were numerous studies of the behaviour
of water soluble polymers near the lower critical solubility
temperature (LCST). Typical systems include
poly-N-isopropylacrylamide (PNIPAM) homopolymer and
block copolymers \cite{Experiment} of the 
poly(ethylene oxide)-poly(propylene oxide)-poly(ethylene oxide),
or PEO-PPO systems briefly \cite{pluronics}.
These copolymers are widely used in pharmaceutical, agricultural
and food industries due to their low toxicity and surfactant
characteristics (EO is hydrophilic and PO is hydrophobic).
Block and more complex random heteropolymers in solutions and melts
at higher concentrations exhibit a variety of
ordered micro--phase separated and disordered glassy phases
\cite{Miles-book,Orland}.
Block copolymers also play role of surfactants in ternary mixtures
of two otherwise immiscible liquids such as water and oil, and
these mixtures also produce micelles and lamellae \cite{Micellization}. 

There are two types of triblock copolymers having rather distinct properties. 
First, a PEO-PPO-PEO polymer with hydrophilic
ends and a hydrophobic middle is the standard Pluronic \cite{Israel,pluronics}, 
which forms stable micelles.
Second, a polymer with the inverse structure would tend to network
hydrophobic end `stickers' by hydrophilic bridges.
These are so--called telechelic associating polymers forming physical
gels with nontrivial rheological properties at higher concentrations
\cite{Telechelic}.

The aggregation number distribution during micellisation has been a 
matter of intensive studies for amphiphilic molecules \cite{Israel}. 
The classical situations, in which there is a clear phase separation 
transition from $M_{cl}=1$ to $M_{cl}=M_{max}$, or where one could introduce 
notions of CMC (critical micelle concentration),
and possibly also CAC (critical aggregation concentration), are well known.
Some cases of molecular self--assembly, in which relatively monodispersed
distributions arise around CMC, have also been seen.
It is a conventional wisdom that in solutions of amphiphilic polymers  
the notion of CMC does not really exist 
since single chain micelles already form in an infinitely dilute 
solution if the primary sequence of the polymer permits micro--phase 
separation within the collapsed globule.

The phase diagrams of diblock and in some cases triblock copolymers 
have been closely investigated \cite{Israel}.
At low concentration these systems 
form micelles, but the question of their aggregation number 
is not so well understood as it appears to be quite sensitive on 
various parameters of the system, although it is generally believed that 
such a distribution should be fairly polydispersed.
This situation is in a marked contrast with a 
two--phase separation onto the macro--aggregate and a gas of single globules 
in the simpler case of homopolymer solutions, which are well described 
by the mean--field Flory--Huggins type theories
\cite{Flory-Huggins}, and their extensions based on 
scaling arguments \cite{Daoud},
self--consistent treatments \cite{Polymer-books} and the Lifshitz approach
\cite{Grosb-92}.

Ordinary micellisation and more complex aggregation number distributions
in copolymer solutions are realised due to the complex interactions 
favouring micro--phase separation of hydrophobic and hydrophilic units. 
Since the CMC of these polymers is essentially zero as
in the case of polysoaps \cite{polysoaps}, one would  like to 
see how the aggregation number of the micelles would behave nearby, 
i.e. in  a very dilute solution. Indeed, normally the character of such a
distribution has a clearly distinct character around CMC vs other regimes.

In attempting to resolve this interesting question, unfortunately, 
both experiment and theory had considerable difficulties and little success. 
The aggregation phenomenon significantly obscures experimental
observation of a single chain collapse by light scattering techniques,
leading often to a diverse range of theoretical
interpretations, particularly as it may be hard to separate purely
equilibrium issues from the kinetic ones \cite{Chu}.
Indeed, the rather skewed shape of the Flory--Huggins diagram requires extremely 
low concentrations to see single globules \cite{WuSingle}
thus making the scattered light
intensity too low for a reliable observation.

In a number of recent experiments on PNIPAM polymers in aqueous solution,
in which some of the units are either hydrophobically modified
or replaced by ionomers \cite{Gorelov,Deng-Qiu}, it has been observed
that stable size monodispersed nanoparticles can be formed on heating
above the LCST instead of chains simply aggregating.
The mean size of these particles is thermodynamically controlled and
it increases with increasing  
polymer concentration and increasing strength of inter--monomer attraction. 
Electron microscopy further confirms the existence of such spherical particles 
with mean size and distribution in agreement with the light scattering.
We called such an unusual type of mesoscopic structures
{\it mesoglobules} as they tend to have
a compact spherical shape and a rather monodisperse size distribution.
To be explicit then their precise definition is as follows.
{\em Mesoglobules are essentially equally sized globules obtained
by association of more than one and less than all polymer chains
in the system} and corresponding to, at least, a local minimum of the free
energy.
Thus, the term mesoglobules refers to the globules possessing a 
well characterised size distribution, but not necessarily a well defined
conformational structure such as e.g. in micelles.

In our previous paper \cite{clusnew} based on the Gaussian variational
theory and lattice Monte Carlo simulation we have shown that in very 
dilute solutions the histograms of cluster distribution favour size 
monodispersed structures for a certain class of heteropolymer sequences 
and only in some narrow strips of the phase diagram.
We also have seen that the mean size of the mesoglobules is related to a 
characteristic scale of the micro--phase separation.
Thus, generally the larger this scale (e.g. block size) the larger are the
mesoglobules produced.

Significantly, in the situation when the mesoglobules occur there is a 
quite narrow single peak distribution in the aggregation numbers and the
mean squared radius of gyration, so that the populations of single 
globules or the macro--aggregate are absent. Moreover, we have demonstrated 
that the mesoglobules are truly thermodynamically stable, i.~e. the 
clusters of exactly equal size possess the lowest free energy value.
Clusters of slightly unequal size obtained from these ideal
mesoglobules have a higher free energy value and the barriers 
separating the mesoglobular minimum from other local minima tend to 
grow quickly as more and more asymmetric divisions of the system into 
macromolecular clusters are considered. 
This situation is quite unusual and novel for spherical micelles and 
it occurs due to a rather delicate competition of energetic
(micro--phase separation) and entropic (translational entropy)
terms under the connectivity constraints.
For the mesoglobules to form the degree of amphiphilicity 
(distinction in the interaction energies between `a' and `b' monomers with 
the solvent) and the mean strength of the monomers attractions should
be in a certain relation. 
In Ref. \cite{clusnew} we have seen that this situation
is possible for alternating monomers, repeating blocks
(having smaller mesoglobules), as well as 
for a number of randomly generated sequences, but in all cases at rather
low concentrations.
Mesoglobules can form at infinitesimally small concentrations,
corresponding to essentially zero CMC in polymeric surfactants.
The mean size of the mesoglobules 
grows with increasing concentration, but when the latter reaches a certain 
value, which perhaps one may view as an analog of CAC,
a polydisperse distribution of clusters with a large population
of the macro--aggregate results, recovering a more conventional regime.

Our understanding of the diblock copolymers, however, was not complete 
in paper \cite{clusnew}. We believed that a finite size effect was
significant there due to the chosen chain lengths.
Thus, in this Letter we shall revisit
the diblocks and also consider the triblocks of both possible types.

\section{Results}


As in Ref. \cite{clusnew}
we adopt the Metropolis technique in the lattice model of 
Ref. \cite{CoplmMonte}.
This model, apart from the connectivity and excluded volume constraints,
includes short--ranged pair--wise interactions between
lattice sites.
The system is completely characterised by three Flory interaction
parameters, $\chi_{aa}$, $\chi_{ab}$ and $\chi_{bb}$,
along with $N$, $M$ and linear lattice size $L$.
In addition to local monomer moves \cite{CoplmMonte}, we
include translational moves representing diffusion of chains.
The latter moves are applied to all clusters
of chains with a probability inversely proportional to
the number of monomers within (Stokes law).
We shall consider sequences consisting of strongly hydrophobic and 
slightly hydrophilic units. However, while in paper \cite{clusnew}
the number of monomers in each chain were $N=24$ and the total
number of chains was $M=20$, here we shall analyse the case of twice the
number $M=40$ of half--long chains $N=12$ in the same simulation box
of linear size $L=60$. Thus, the monomer concentrations $c\equiv NM/L^3$ are
the same in both cases, and also we shall keep the numbers of the hydrophobic
and hydrophilic monomers equal in each chain.

In Figs.~\ref{fig:snaps} we present snapshots of two typical
types of structures observed for triblock copolymers in the region
of the phase diagram where mesoglobules are normally seen \cite{clusnew}.
Fig.~\ref{fig:snaps}a corresponds to triblocks with the hydrophilic (or
P- for polar) ends and the hydrophobic (or H-) middle block, and we shall
call these the P-ends triblocks briefly. These structures clearly have
essentially equal size and thus could be considered as mesoglobules.
Moreover, they possess a pronounced micellar character with a hydrophobic
(black circles) core and a hydrophilic (white circles) corona.
Fig.~\ref{fig:snaps}b corresponds to the inverted case of the H-ends triblocks.
In this case typical structures often have two, three (as in this figure), or
more large clusters inter--connected by the hydrophilic middle block
bridges. Such dimers, trimers and multimers are indeed to be expected
at very low concentrations as it is known that this type of associating
polymers forms telechelic physical gels at higher concentrations 
\cite{Telechelic}.

To investigate the aggregation number and size distributions 
we have obtained a large ensemble of $Q=1000$ independent equilibrium states. 
In  Figs.~\ref{fig:hysto} we present the calculated
histograms of the aggregation number (mass) and size probability
densities for the diblock ($a_6 b_6$) and triblock sequences (H-ends
and P-ends).
We have seen in Fig.~7b of Ref. \cite{clusnew} that the diblock 
sequence s4 had a large population of the macro--aggregate there.
We have argued in Ref. \cite{clusnew} that this might be due to 
a finite--size effect since the
total number of chains $M=20$ was comparable to the mean aggregation
number of the mesoglobules.
In Fig.~\ref{fig:hysto}a, however, one can see for the diblocks that
their mean aggregation
number in the mesoglobules is about $12$, whereas the total number of
chains is $M=40$, so that there is no population of the macro--aggregate
any more. Therefore, the diblock sequences do indeed produce
mesoglobules in a certain region of the phase diagram in dilute solution
and the corresponding size distribution in Fig.~\ref{fig:hysto}b is quite
narrow.

Similarly to diblocks, the P-ends triblocks produce a clear
single peak distribution for the aggregation number (see Fig.~\ref{fig:hysto}a) 
corresponding to the mesoglobules of a somewhat smaller size and 
the respective cluster size distribution is also rather narrow 
in Fig.~\ref{fig:hysto}b.
On the contrary, the aggregation number distribution for the H-ends triblocks 
is rather broad and has a certain population of the macro--aggregate.
The width of this distribution is explained mainly by formation of
dimer and multimer clusters, so that one can no longer define a preferred
cluster size. Fig.~\ref{fig:hysto}b shows that the size distribution
of the H-ends has two peaks: a very broad multimeric peak 
(near the region $6<R_{cl}<8$) and a still reasonably
narrow peak corresponding to the mesoglobules and their dimers
(near $R_{cl}\approx 4$). It is interesting to note that the
mean size of the latter objects for the H-ends triblocks is
considerably smaller than for the diblock sequence of the same
length, although the situation is reverse for the mean aggregation
numbers. This could be easily understood by comparison of
Figs. 6d in Ref. \cite{clusnew} and ~\ref{fig:snaps}b. One could see 
that the P-corona in the former case is quite extended, whereas in the
latter case it is more dense and compact, containing a considerable
number of loops.

Finally, it is also instructive to compare the mesoglobules
for diblock $a_6 b_6$ with those of twice longer sequence
$(a_6 b_6)_2$ (see s3 in Figs. 7a,8a in Ref. \cite{clusnew})
formed by merging every two $a_6 b_6$ chains  at the same monomer 
concentration $c$. By a careful analysis of these data
one can conclude that the mean cluster mass increases
and the mean size decreases slightly for $(a_6 b_6)_2$ as
compared to $a_6 b_6$. The effect of an increase in the mean cluster mass
is rather weak and does not have a simple explanation, but the decrease
of the mean size is clearly due to a more compact size of loops
in mesoglobules of $(a_6 b_6)_2$ than the size of free P-ends
for $a_6 b_6$, similarly to the discussion in the previous
paragraph.


\section{Conclusion}

In this Letter we have addressed the question of sequence specificity 
for forming mesoglobules in dilute solutions. The aggregation number and 
size distributions for inter--molecular clusters in a poor solvent have 
been found to be monodisperse for diblocks and triblocks with outer 
hydrophilic ends, while rather polydisperse for triblocks with 
hydrophobic end 'stickers'. Comparative analysis of chains of varying 
length but at a fixed total monomer concentration has elucidated  
nontrivial dependences of the mesoglobules mass and size on the copolymer 
chain length and sequence and a special role of hydrophilic loops and 
bridges in determining the compactness of the resulting structures. 

The main conclusion from these studies is that the mesoglobules formation 
thermodynamics is rather delicate and that such stable structures are 
only possible for certain heteropolymer sequences and in a narrow region 
of the phase diagram.
Our results also suggest that multimeric clusters of a certain 
size may have a stabilisation mechanism similar to that of mesoglobules, 
and again only for certain heteropolymer sequences.

We believe that the observations of this Letter may shed some light on
understanding the problems of competition aggregation vs folding 
in protein solutions and, possibly, on 
self--organisation of the quaternary structure in
multimeric proteins. It appears that for the latter
process to take place a considerable number of `sticky'  
hydrophobic amino acid residues should be exposed on the exterior of 
each of folded subunits.
The resulting association produces a well defined quaternary structure,
which is biologically functional as in e.g. hemoglobin,
rather than a disordered aggregate. 
Hopefully, work in this direction would also help unravel
the mechanism of association of so--called Bence--Jones proteins
believed to be related to amyloid fibril formation.

More importantly,
detailed understanding of the mechanism of mesoglobules formation
is important for being able to select polymers from which nanoparticles 
of required size and polydispersity can be prepared. 
The main advantage of this new approach is that nanoparticles 
can be formed reversibly in mild conditions from a prepurified polymer. 
It also seems quite reasonable that even a weak electrostatic repulsion
may play a crucial role for further stabilisation of mesoglobules
and improving their monodispersity \cite{Deng-Qiu}.
Nanoparticles carrying functional groups properly located
in the mesoglobule can be obtained, which could be further modified. 
Such nanoparticles with well controlled size distribution
can find a broad range of applications in pharmaceutical, 
biotechnological and cosmetic industries, where the technological
control is one of the major problems.


\acknowledgments

The authors acknowledge interesting discussions with
Professor H.~Orland, Professor T.~Garel
and our colleague Dr A.V.~Gorelov.
This work was supported by grants SC/99/186 and FR/2000/019
from Enterprise Ireland.

\begin{figure}
\caption{ 
Snapshots of typical conformations from simulation for different sequences.
Fig.~a corresponds to the P-ends sequence $b_3 a_6 b_3$ for which
mesoglobules are formed.
Fig.~b corresponds to the H-ends sequence $a_3 b_6 a_3$, where
networks of locally collapsed clusters connected by hydrophilic bridges
are formed.
Here the equilibration time was 
$1.92 \cdot 10^9$ of attempted Monte Carlo moves and other
parameters are: $L=60$, $N=12$,
$M=40$, $\chi_{aa}=1$, $\chi_{ab}=0.4$ and $\chi_{bb}=-0.2$.
Black (white) circles correspond to hydrophobic (hydrophilic)
monomers.
}\label{fig:snaps}
\end{figure}

\begin{figure}
\caption{ 
Probability densities of the number of chains in a cluster
(aggregation number), $M_{cl}$,
(Fig.~a) and of the radius of gyration of a cluster (size), $R_{cl}$,
(Fig.~b) for diblock (thin dashed curve) and triblocks
(thick dashed and solid lines) sequences.
These results have been obtained by analyzing data for
the ensemble size $Q=1000$. Other parameters are as in Fig.~\ref{fig:snaps}.
}\label{fig:hysto}
\end{figure}

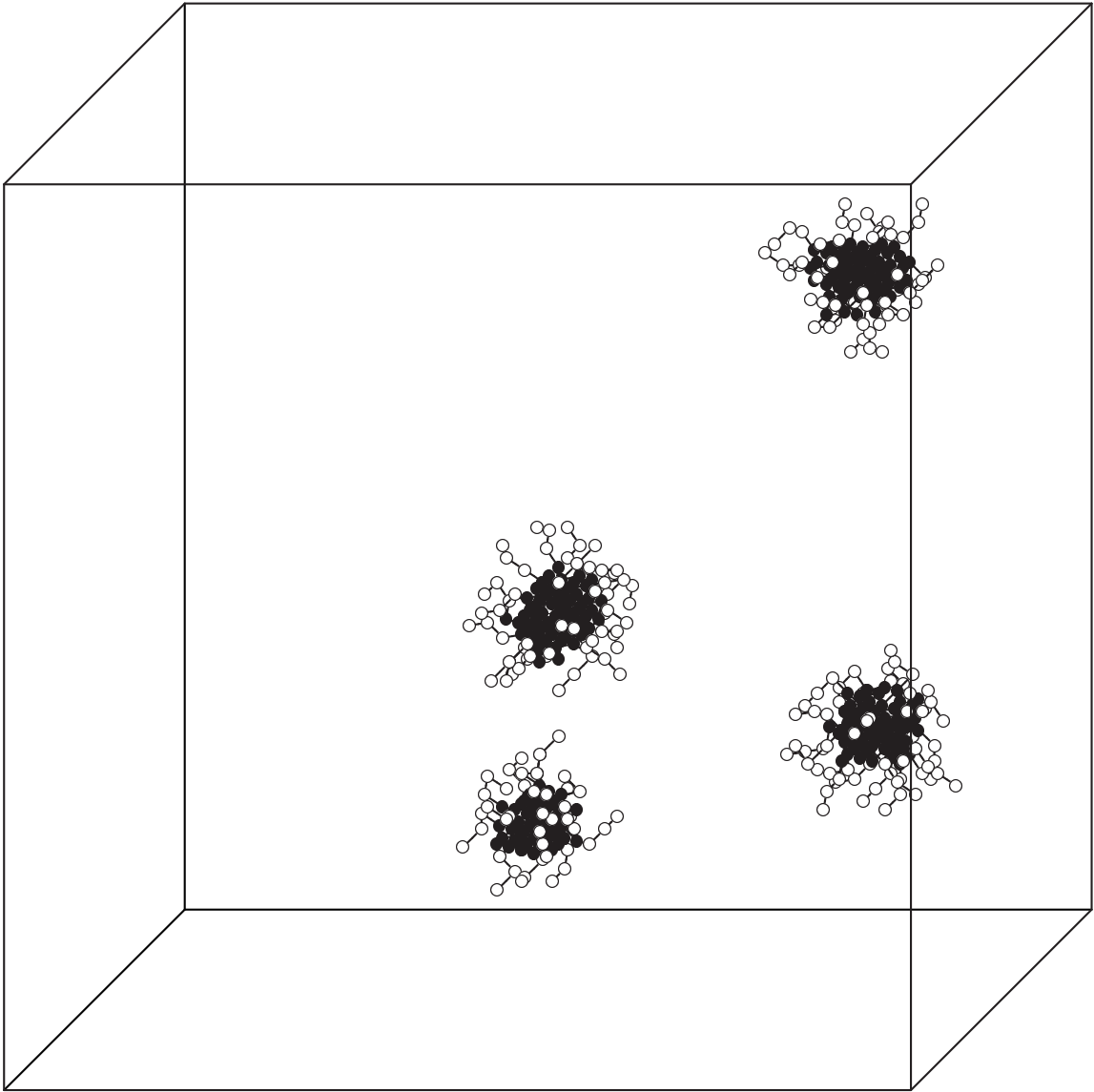
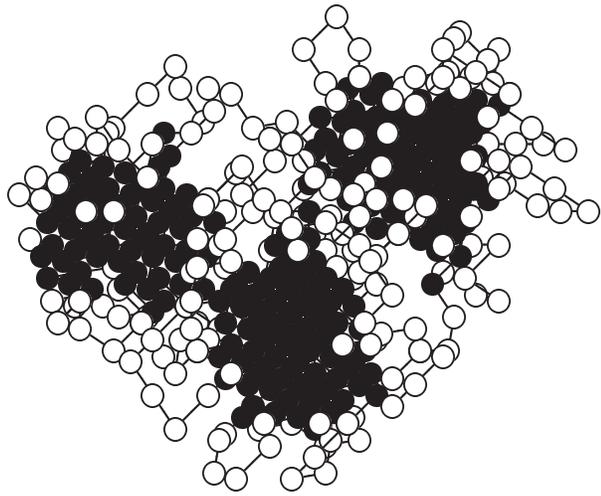
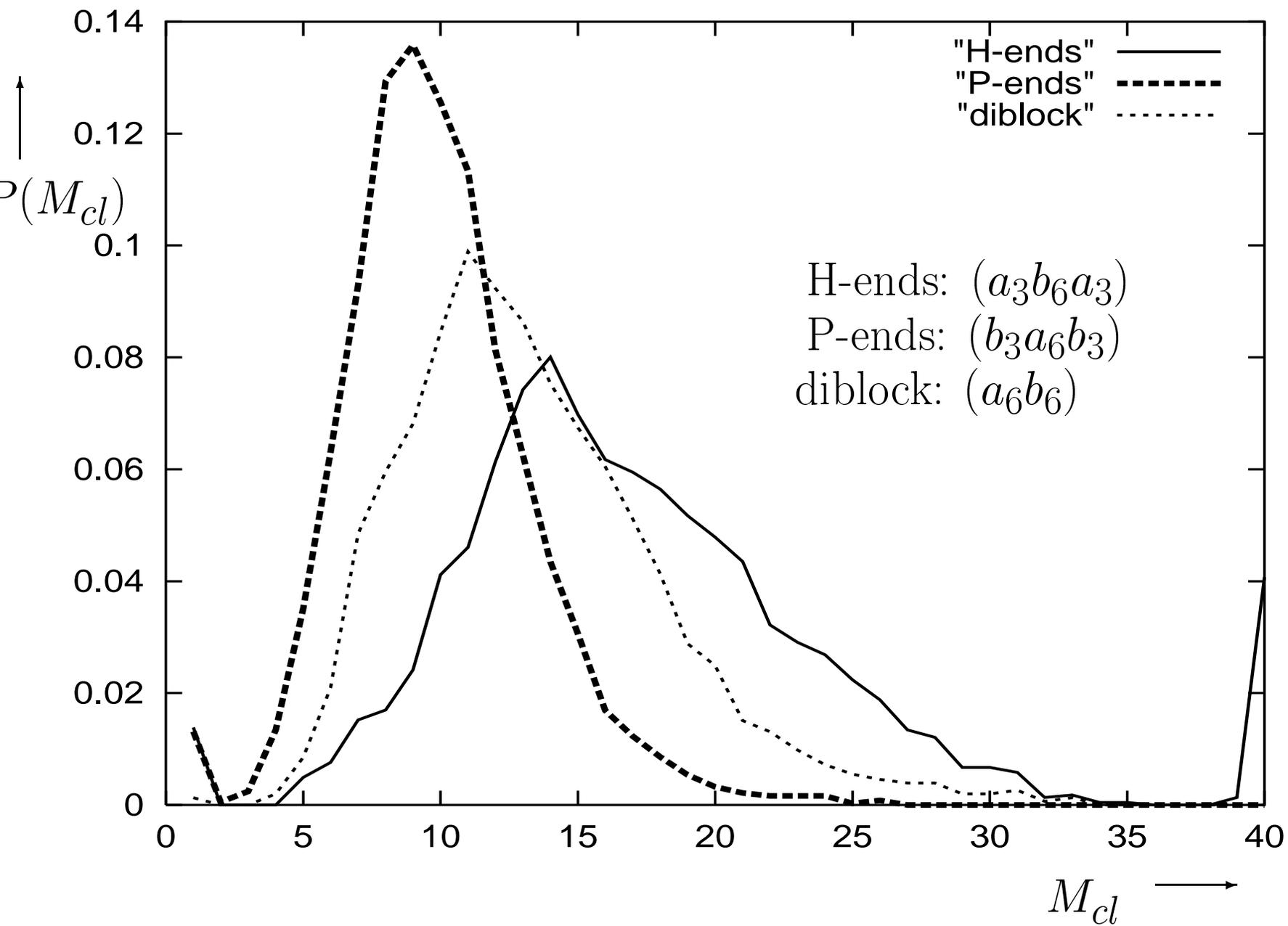
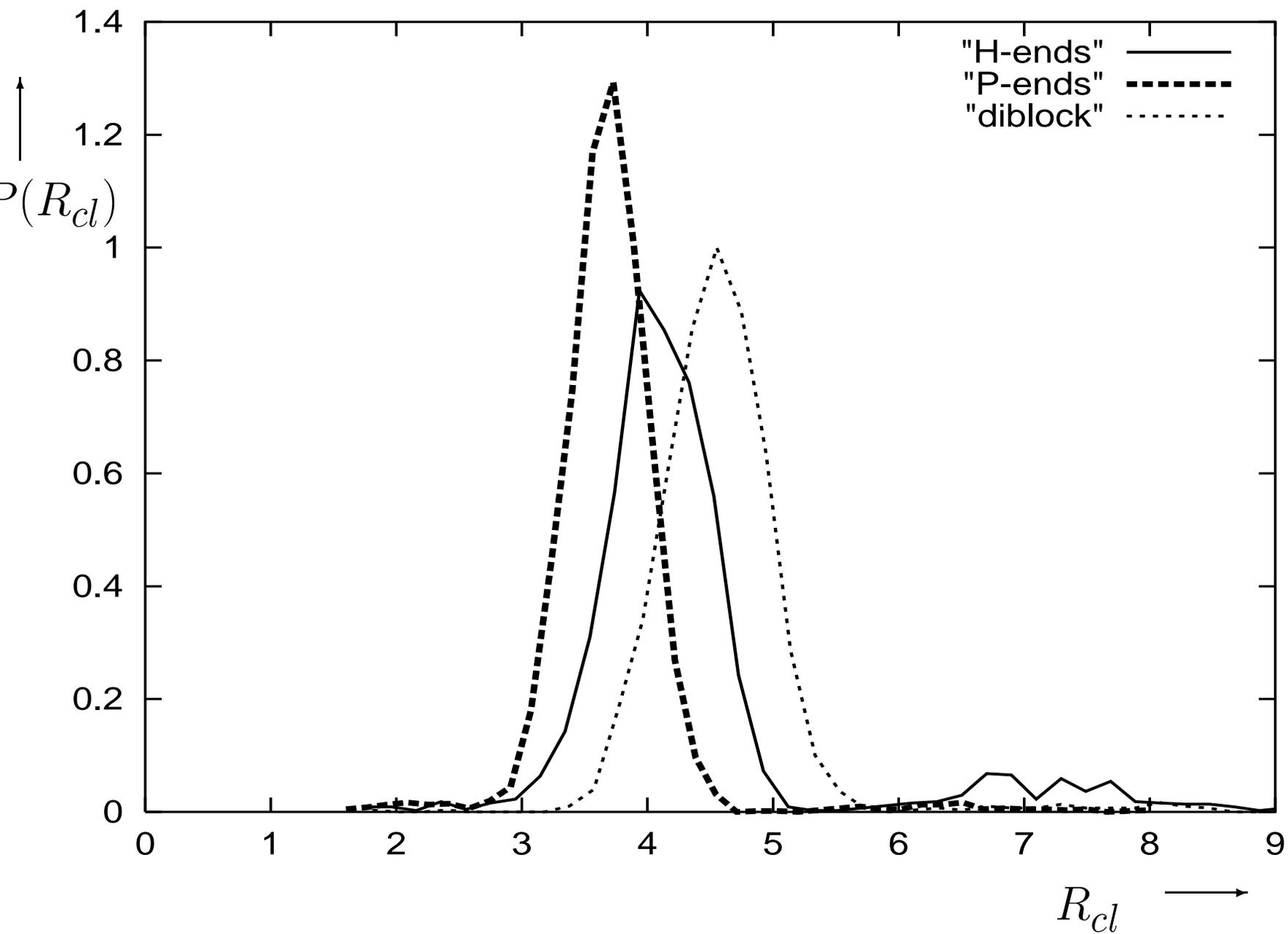
\end{document}